\documentclass[aps,nofootinbib,groupedaddress]{revtex4}

\usepackage{url}
\usepackage{graphicx}
\usepackage{amsmath,amssymb,amstext,amssymb,amsfonts,amsthm}
\usepackage{hyperref}
\usepackage{rotating}
\usepackage[margin=1.0in,papersize={8.5in,11in}]{geometry}
\usepackage[utf8]{inputenc}

\begin{document}

\title{Simulated reconstruction of the remote dipole field using the kinetic Sunyaev
Zel'dovich effect}

\author{Juan I. Cayuso${}^{1, 2}$}
\email[]{jcayuso@perimeterinstitute.ca}
\author{Matthew C. Johnson${}^{1, 3}$}
\email[]{mjohnson@perimeterinstitute.ca}
\author{James B. Mertens${}^{1, 3, 4}$}
\email[]{mertens@yorku.ca}

\affiliation{${}^1$Perimeter Institute for Theoretical Physics, Waterloo, Ontario N2L 2Y5, Canada}
\affiliation{${}^2$Department of Physics and Astronomy, University of Waterloo, Waterloo, ON, N2L 3G1, Canada}
\affiliation{${}^3$Department of Physics and Astronomy, York University, Toronto, Ontario, M3J 1P3, Canada}
\affiliation{${}^4$Canadian Institute for Theoretical Astrophysics, University of Toronto, Toronto, ON M5H 3H8 Canada}

\begin{abstract}
The kinetic Sunyaev Zel'dovich (kSZ) effect, cosmic microwave background (CMB) anisotropies induced by scattering from free electrons in bulk motion, is a primary target of future CMB experiments. In addition to shedding light on the distribution of baryons and the details of the epoch of reionization, measurements of the kSZ effect have the potential to address fundamental questions about the structure and evolution of our Universe on the largest scales and at the earliest times. This potential is unlocked by combining measurements of small-scale CMB anisotropies with large-scale structure surveys, a technique known as kSZ tomography. Previous work established a quadratic estimator for the remote dipole field, the CMB dipole observed at different locations in the Universe, given a CMB map and a redshift-binned map of large scale structure. This previous work did not include gravitational lensing, redshift space distortions, or non-linear evolution of structure. In this paper, we investigate how well the remote dipole field can be reconstructed in the presence of such effects by using mock data from a suite of simulations of gigaparsec-sized regions of the Universe. To properly model both large and small scales, we develop a novel box-in-box simulation pipeline, 
where small-scale information is obtained from L-PICOLA N-body simulations, and large-scale information obtained by evolving fields using linear theory and adding the resulting corrections to the N-body particle data. This pipeline allows us to create properly correlated maps of the primary CMB including lensing as well as the kSZ effect and density maps on the past light cone of an observer. Analyzing an ensemble of mocks, we find that the dipole field can be reconstructed with high fidelity over a range of angular scales and redshift bins. However, we present evidence for a bias due to the non-linear evolution of structure. We also analyze correlations with the primary CMB, investigating the ability of kSZ tomography to reconstruct the intrinsic CMB dipole. Our results constitute a proof-of-principle that kSZ tomography is a promising technique for future datasets.
\end{abstract}

\maketitle

\section{Introduction}

Fueled by rapid improvement on the experimental and data analysis fronts,
our theoretical understanding of the Universe has condensed into the standard
model of cosmology, $\Lambda$CDM. This model is able to
describe our Universe with a high degree of accuracy. Nevertheless, the fundamental
nature of the major constituents of the model remains unknown, and a number of 
potential anomalies remain unexplained  (see \cite{1510.07929} for a recent review).
The primary cosmic microwave background (CMB) has thus far been the workhorse of cosmology,
however the primary CMB temperature anisotropies have now been measured across an impressive
range of angular scales to their ultimate cosmic variance limit by the Planck satellite.
While additional progress can be made using measurements of CMB polarization, it will be
necessary to cultivate additional observables to further improve.

One of the next frontiers of observational cosmology lies in the secondary CMB, temperature and polarization anisotropies induced by the scattering of CMB photons from mass (lensing) or free charges (the Sunyaev Zel'dovich effect). These effects are important on angular scales of roughly an arcminute, where the power in the primary CMB is rapidly falling due to Silk damping. In this paper, our primary focus is the kinetic Sunyaev Zel'dovich (kSZ) effect, temperature anisotropies generated by CMB photons scattering off of free electrons in bulk motion. The kSZ effect is the dominant blackbody contribution to the CMB on small angular scales. Although the amplitude of fluctuations is small, of order a microkelvin, the kSZ effect has now been detected at greater than $4\sigma$ \cite{1203.4219, 1607.02139, 1603.03904, 1504.03339}, with future experiments~\cite{1610.02743} forecasted to achieve signal to noise in excess of $10^2$. 

The kSZ effect can be expressed as the line of sight integral~\cite{Sunyaev:1980nv}
\begin{equation}\label{eq1}
\frac{\Delta T}{T}\Bigg{|}_{kSZ}(\hat{\textbf{n}}_{e})  =-\sigma_{T}\int_{0}^{\chi_{re}}d\chi_{e}a_{e}
(\chi_{e})\bar{n}_{e}(\chi_{e})(1+\delta_e(\hat{\textbf{n}}_{e},\chi_{e}) )v_{\mbox{\footnotesize{eff}}}
(\hat{\textbf{n}}_{e},\chi_{e}),
\end{equation}
where $\sigma_{T}$ is the Thompson cross-section, $\bar{n}_{e}(\chi_{e})$ is the average 
electron number density at comoving distance $\chi_{e}$, $\delta_e(\hat{\textbf{n}}_{e},
\chi_{e})$ is the electron overdensity field, $\hat{\textbf{n}}_{e}$ is the angular direction on 
the sky to the electron, $v_{\mbox{\footnotesize{eff}}}(\hat{\textbf{n}}_{e},\chi_{e})$ is 
the projection of the remote CMB dipole field (the CMB dipole observed by each electron along the line of sight), 
and $a_{e}$ is the scale factor at  the scatterer's location.

The remote dipole field at any given point in spacetime depends on the local peculiar velocity of electrons as well as primordial contributions from the surface of last scattering 
(SLS). Because each electron samples a different portion of the SLS, the kSZ effect in principle contains more information about fluctuations on the largest scales than the primary CMB alone~\cite{astro-ph/9703118, 0711.3459, 1009.3967, 1111.3794, 1104.1300, 1105.0909, 1108.2222, 1008.0469, Zhang:2015uta, 1610.06919, 1707.08129}. However, much of this extra information is diluted by the line of sight integral in Eq.~\ref{eq1}, presenting a challenge for making progress with the kSZ power spectrum alone. 

Key to extracting information about the dipole field, and therefore the largest scales, is to use cross-correlations with probes of large-scale structure (LSS) such as galaxy surveys and 21 cm measurements, a technique known as kSZ tomography~\cite{Ho:2009iw,1004.0990,1004.1301}. While a number of variants exist, we focus on direct cross-correlations between the small angular scale CMB and the redshift-binned density field. A set of theoretical tools for kSZ tomography was developed in~\cite{1610.06919}, and an optimal quadratic estimator derived in~\cite{1707.08129}. Schematically, the correlation function is $\langle \Delta T_{\rm kSZ} \ \delta \rangle \sim \langle v_{\rm eff} \delta \delta_e \rangle \sim v_{\rm eff} \langle \delta \delta_e \rangle$ where $\delta$ is the overdensity field for the tracer.
Importantly, since the dipole field receives contributions mainly from large scale modes, while the density fields receive contributions mainly from small-scale modes, the result is an isotropic power modulated by the dipole field.
In analogy with reconstruction techniques for CMB lensing~\cite{astro-ph/0301031} and patchy reionization~\cite{0812.1566}, this statistical anisotropy is the basis for reconstructing the remote dipole field. By constructing correlators for each redshift bin, one can reconstruct the fully three-dimensional coarse-grained dipole field. Ref.~\cite{1707.08129} forecasted that a high fidelity reconstruction of the dipole field should be possible with next-generation galaxy surveys such as LSST~\cite{0912.0201} and next-generation CMB experiments such as CMB-S4~\cite{1610.02743}. 
 
The goal of this paper is to further explore the reconstruction of the remote dipole field by analyzing a set of mock CMB and LSS maps generated from a suite of  simulations. Such simulations allow us to explore previously neglected effects such as gravitational non-linearities, redshift space distortions, and CMB lensing. However, a proper treatment of all relevant physics is intrinsically challenging.
One must model both the dipole field, which receives important contributions from scales of order the size of the observable Universe, as well as the density fields, which depend on small scales and include baryonic physics. Evolving a standard N-body simulation incorporating such a large range of scales is currently computationally intractable. To overcome this limitation, we develop a novel box-in-box simulation framework which consistently embeds a $\sim$Gpc-sized N-body simulation inside of a box whose volume is larger than the observable Universe, and which contains large-scale modes evolved using linear theory.
This box-in-box procedure allows us to use the data from both of these simulations to produce properly correlated maps of the lensed primary CMB temperature anisotropies, kSZ temperature anisotropies, the dipole field, and the dark matter overdensity field. We do not model baryonic physics in the present analysis, and therefore use the dark matter density as a proxy for the electron density. However, because our simulation framework is modular, N-body simulations including baryonic physics will be incorporated in the future.

We find that the quadratic estimator efficiently reconstructs the remote dipole field over a range of angular scales and redshift bins, indicating that kSZ tomography is generally robust. However, we present evidence for a bias due to non-linear structure at low redshifts. We demonstrate the ability of kSZ tomography to reconstruct the fundamental component of the observed CMB dipole, supporting the suggestion in previous work~\cite{1707.08129} that this could be an early application of kSZ tomography on large angular scales. The results we present here are intended primarily as a proof-of-principle both of the simulation framework and remote dipole reconstruction. To lay the groundwork for the analysis of near-term datasets, various layers of realism will be added to our simulation framework in future work, including the construction of mock galaxy catalogs, improved resolution, inclusion of baryonic physics, correlated foregrounds such as thermal SZ, and partial sky data. 

\section{The remote dipole field}\label{sec:theory}

In this section we briefly present expressions for the remote dipole field, its theoretical power spectrum, and the real-space quadratic estimator
that we use to reconstruct the dipole field.
Further details on the velocity power spectra, transfer functions and harmonic-space quadratic estimators can be found 
in \cite{1610.06919, 1707.08129}. In order to work with a binned version of Eq.~\ref{eq1}, we consider a bin-averaged remote dipole field  
$\bar{v}^{\alpha}_{\mbox{\footnotesize{eff}}}(\hat{\textbf{n}}_{e})$, which
 can be expressed in terms of contributions to the CMB temperature
$\Theta(\hat{\textbf{n}}_{e},\chi_{e}, \hat{\textbf{n}} )$ 
seen along the 
sky direction $ \hat{\textbf{n}}$ by free falling electrons at positions $\textbf{r}_{e}=\chi_{e}\hat{\textbf{n}}_{e}$ inside each redshift bin:
\begin{equation}\label{eq3}
\bar{v}^{\alpha}_{\mbox{\footnotesize{eff}}}(\hat{\textbf{n}}_{e})=\frac{3}{4\pi}\frac{1}{\Delta\chi_{\alpha}}\int_{\chi^{\alpha}_{min}}^{\chi^{\alpha}_{\max}}d\chi_{e}\int 
d^{2}\hat{\textbf{n}}\:\Theta(\hat{\textbf{n}}_{e},\chi_{e}, \hat{\textbf{n}} )\:(
\hat{\textbf{n}}\cdot\hat{\textbf{n}}_{e}),
\end{equation}
Here, the index $\alpha$ labels each bin, which extend over the range in comoving distance $\chi^{\alpha}_{min} < \chi <\chi^{\alpha}_{max}$, and where $\Delta\chi_{\alpha} = \chi^{\alpha}_{max} - \chi^{\alpha}_{min}$. The radiation field is 
\begin{equation}\label{eq4}
\Theta(\hat{\textbf{n}}_{e},\chi_{e}, \hat{\textbf{n}} )= \Theta_{\mbox{\footnotesize{SW}}}
(\hat{\textbf{n}}_{e},\chi_{e}, \hat{\textbf{n}} )+
\Theta_{\mbox{\footnotesize{ISW}}}(\hat{\textbf{n}}_{e},\chi_{e}, \hat{\textbf{n}} )+
\Theta_{\mbox{\footnotesize{Doppler}}}(\hat{\textbf{n}}_{e},\chi_{e}, \hat{\textbf{n}} ),
\end{equation}
which receives contributions from the Sachs-Wolfe
effect, the integrated Sachs-Wolfe effect due to the evolution of the gravitational
potential along the line of sight, and the Doppler effect due to peculiar motion of electrons at $\textbf{r}_{e}$ relative 
to the SLS (see e.g.~\cite{Dodelson:2003ft}). The binned power spectrum is given by
\begin{equation}\label{CL_vv}
C_{\alpha\beta l}^{\bar{v}\bar{v}} = \int\frac{d^{3}k}{(2\pi)^{3}}P_{\Psi}(k){\Delta}^{\bar{v}*}_{\alpha l}(k){\Delta}^{\bar{v}}_{\beta l}(k),
\end{equation}
where Greek indices denote redshift bins, $P_{\Psi}(k)$ is the power spectrum of the Newtonian gauge primordial gravitational 
potential $\Psi$, and ${\Delta}^{\bar{v}}_{l}(k,\chi_{e})$ is the remote 
dipole transfer function, given in Ref.~\cite{1707.08129}. 
As shown in \cite{1707.08129}, the presence of a 
large scale dipole will manifest in the cross correlation between the kSZ contribution to the CMB temperature
and the moments of a redshift binned density distribution $\delta^{\alpha}$ defined by
\begin{equation}\label{eq5}
\delta^{\alpha}(\hat{\textbf{n}})= \frac{1}{\Delta\chi_{\alpha}}\int_{\chi^{\alpha}_{min}}^{\chi^{\alpha}_{\max}}d\chi\;\delta(\hat{\textbf{n}},\chi).
\end{equation}

A real-space optimal quadratic estimator for the moments of the bin-averaged remote dipole field, $\widehat{v}_{\mbox{\footnotesize{eff}},lm}^{\alpha}$, is given by: 
\begin{equation}\label{eq6}
\widehat{v}_{{\rm eff}, \ell m}^{\alpha} =  {N^{\bar{v}\bar{v}}_{\alpha \ell}} \int d^2 \hat{n} \ Y_{\ell m}^* (\hat{n}) \xi(\hat{n}) \zeta^\alpha (\hat{n}).
\end{equation}
\begin{eqnarray}
\label{eq:xi}
\xi(\hat{n}) &=& \sum_{\ell m} \frac{a_{\ell m}^T}{{C}_\ell^{TT}} Y_{\ell m}(\hat{n}) \\
\label{eq:zeta}
\zeta^\alpha (\hat{n}) &=& \sum_{\ell m} \frac{\delta^{\alpha}_{\ell m}{C}^{\delta \tau}_{\alpha \ell}}{{C}^{\delta \delta}_{\alpha \ell}} Y_{\ell m}(\hat{n})
\end{eqnarray}
where
\begin{equation}\label{eq7}
\frac{1}{N^{\bar{v}\bar{v}}_{\alpha l}} = \frac{1}{2l+1}\sum_{l_{1}l_{2}}\frac{\Gamma^{\mbox{\tiny{\;kSZ}}}_{l_{1}l_{2}l \alpha}
\Gamma^{\mbox{\tiny{\;kSZ}}}_{l_{1}l_{2}l \alpha}}{C^{TT}_{l1}C^{\delta\delta}_{\alpha l_{2}}}.
\end{equation}

The coupling constant $\Gamma^{\mbox{\tiny{\;kSZ}}}_{l_{1}l_{2}l \alpha}$ is defined by
\begin{equation}\label{eq8}
\Gamma^{\mbox{\tiny{\;kSZ}}}_{l_{1}l_{2}l \alpha} = \sqrt{\frac{(2l_{1}+1)(2l_{2}+1)(2l+1)}{4\pi}}
\begin{pmatrix} l_{1} & l_{2} & l \\ 0 & 0 & 0 \end{pmatrix}C^{\tau\delta}_{\alpha, l_{2}},
\end{equation}
where the quantities with parenthesis are Wigner 3j symbols and $C^{\tau\delta}_{\alpha, l_{2}}$ is the cross-power
between the binned galaxy density and the anisotropies in the optical depth of the redshift bin
\begin{equation}\label{eq9}
\tau^{\alpha}(\hat{\textbf{n}}) = -\sigma_{T}\int_{\chi_{min}^{\alpha}}^{\chi_{max}^{\alpha}}d\chi\:a(\chi)\:\bar{n}_{e}(\chi)(1+\delta_e(\hat{\textbf{n}},\chi)).
\end{equation}
Because the simulations presented below do not contain baryons, we assume that the electron density field traces the dark matter density field.

\section{Simulations}
\label{sec:sims}

Our simulation framework includes two components: a small-scale N-body
simulation and a large-scale random field evolved
using linear perturbation theory. We explore the idea of ``sewing'' these
simulations together in order to accurately model physics on both
large and small scales, thereby obtaining consistent realizations
of both the primary CMB and angular, projected matter fields.

In order to obtain lightcone data on small scales, we use the publicly
available L-PICOLA code \cite{1506.03737}. L-PICOLA is a ``Lightcone-enabled
Parallel Implementation of the COLA'' method, providing an efficient
means for generating both data on an observer's past lightcone and 
data on spatial hypersurfaces. The COmoving Lagrangian
Acceleration (``COLA'') method \cite{1108.5512, 1301.0322} works by solving
the second-order Lagrangian perturbation theory (2LPT) equations in
order to generate an initial guess for the motion of particles in
the simulation, and subsequently solves a set of equations describing
the difference between the 2LPT solution and the full N-body equations
in order to improve the accuracy of the 2LPT solution. This method
allows L-PICOLA to obtain results with an accuracy similar to full
N-body simulations on the scales we are interested in, but with a
substantially larger simulation timestep, and therefore at a substantially
reduced computational cost. In the limit of many timesteps, the output
from L-PICOLA should be equivalent to a traditional N-body simulation.

Although these N-body simulations are able to provide us with particular
realizations of physics on small-scales, we are interested in modeling
both the primary CMB and kSZ temperature fields. In order to obtain
contributions to the kSZ temperature from the full dipole field 
(Eq.~\ref{eq3}) in a manner consistent with the small-scale L-PICOLA
data, as well as to generate the primary CMB, we utilize
a novel ``box-in-box'' technique. This technique is similar in spirit
to the mode-adding procedure (MAP) described in
\cite{astro-ph/9512131, astro-ph/9604046}, in that information about
large scales is added to a small-scale simulation.
However, the technique we utilize differs in several important regards.
Similar to \cite{astro-ph/0103301}, we add information at the level of
the density and peculiar velocity fields directly in Eulerian or configuration
space, rather than in either Fourier space or Lagrangian space;
additionally, no information is removed from the small-scale simulation.

We utilize N-body simulations with a number of particles
$N_p = 1280^3$ in a comoving volume $(2 {\rm Gpc/h})^3$, corresponding to a
maximum simulation redshift of $z \sim 0.37$ and particle mass 
$2 \times 10^{12}\,{\rm M}_\odot$. While this coarse resolution does not allow
us to resolve the structure of small mass halos, and also does not necessarily result
in high-fidelity simulation data on the associated length scales, we find that the
data we do obtain is sufficient for use in producing maps at angular resolutions
of interest to us. We require the large-scale
random field to encompass a volume containing the CMB (and ideally larger modes),
so we utilize a large-scale ``box'' with volume $(32 {\rm Gpc/h})^3$, resolved by
$320^3$ grid points.

The ``box-in-box'' method should be valid in a regime similar
to the MAP method, which itself has been shown to perform
well when linear theory provides a good description of the field
content. This is a somewhat stronger condition than requiring mode
amplitudes or the power spectrum to be well-described
by linear theory. While the linear and nonlinear matter power spectra
agree to within a few percent down to scales of order 10 Mpc, mode
coupling can exist -- nonlinear terms of order $\delta \rho^2$ can
constitute percent or larger corrections to evolution of the density
field on scales of order a few hundred Mpc. So long as we remain in
a regime where the field configuration is sufficiently well-described
by (only) linear theory, we can expect the box-in-box technique
to work. For the Gpc-scale N-body box sizes we employ here,
this is the case.

\subsection{Simulating small scales using L-PICOLA}

We make use of both the lightcone output from L-PICOLA as well as
data from spatial slices. The particle data from spatial slices is used
to compute both primordial and large-scale components of the kSZ
and primary CMB, and will be discussed in Section~\ref{subsec:Large-scales}.
The lightcone data is used to construct lightcone-projected sky maps
of the density contrast field, velocity, and momentum fields, as well
as convergence maps. 

We generate radially binned maps of
various fields, both in order to examine the underlying physics of
the simulations, as well as to test reconstruction techniques at various redshifts. We
divide the lightcone data into a number of radial bins between
us and the largest redshift probed by the simulation. These radial bins can then be selectively
integrated over to construct the contributions to a given field,
such as density or kSZ temperature, from a given redshift range.

In order to produce density maps in both radial and angular bins, we bin particle
data by noting that 
\begin{equation}
\delta_{{\rm bin}}=\frac{\rho_{{\rm bin}}-\bar{\rho}_{{\rm bin}}}{\bar{\rho}_{{\rm bin}}}=\frac{n_{{\rm bin}}}{\bar{n}_{{\rm bin}}}-1\,,
\end{equation}
where $\rho$ is the physical density inside a radial-angular-bin on the lightcone
with comoving volume $V_{{\rm bin}}=\frac{\Omega_{{\rm bin}}}{3}(\chi_{B}^{3}-\chi_{A}^{3})$,
where the bin has radial boundaries at $\chi_{A}$ and $\chi_{B}$,
and subtends a solid angle $\Omega_{{\rm bin}}$. The number of simulated
particles of mass $m$ in a bin is $n=\rho/m$, and the average/expected/background
number of particles in a pixel is
\begin{align*}
\bar{n}_{{\rm bin}} & =N_{{\rm sim}}\frac{V_{{\rm bin}}}{V_{{\rm sim}}}=\frac{N_{{\rm sim}}}{V_{{\rm sim}}}\frac{\Omega_{{\rm bin}}}{3}(\chi_{B}^{3}-\chi_{A}^{3})
\end{align*}
in the case of discrete bins, or
\[
\bar{n}_{{\rm bin}}=\frac{N_{{\rm sim}}}{V_{{\rm sim}}}\frac{\Omega_{{\rm bin}}}{3}3\chi^{2}d\chi
\]
in the continuum limit, with $N_{{\rm sim}}$ the total number of
particles in a simulation of comoving volume $V_{{\rm sim}}$. The
overdensity is then given by taking $n_{{\rm bin}}$ to be the number
of particles in a given bin, so explicitly,
\begin{equation}
\label{eq:density_bin}
\delta_{{\rm bin}}=-1+\sum_{{\rm particles}\in{\rm bin}}\frac{1}{\bar{n}_{{\rm bin}}}\,.
\end{equation}
This expression is similar in spirit to that of \cite{1801.05745}, although
not identical. We also integrate the density contrast
along a line of sight \textendash{} or in a pixel subtending some
solid angle on the sky; this can be written as a sum over the densities
of all bins along the line of sight of the pixel, 
\begin{align}
\delta & =\int d\chi\delta(\chi)=\sum_{{\rm bin}\in{\rm pix}}\delta_{{\rm bin}}d\chi_{{\rm bin}}\,,
\end{align}
where the bins that lie along the direction of the pixel on the sky are
summed over. For different choices of radial binning, the sum will
agree up to terms $\mathcal{O}(d\chi^{2})$.

We are additionally interested in accounting for redshift-space distortions (RSDs)
within this framework, requiring a small modification to the density field
used in the reconstruction, Eq.~\ref{eq5}. In order to take RSDs into account, we
perturb particle positions by a small amount corresponding to the
mis-inferred distance. Quantitatively, we compute
\begin{equation}
\chi_{\rm RSD} = \chi_{\rm FRW} ( z_{\rm FRW}(\chi) + v_{\rm Doppler} )\,,
\end{equation}
where functions with the FRW subscripts indicate the background FRW cosmology
has been used, and where we then bin particles using Eq.~\ref{eq:density_bin}
but according to their position $\chi_{\rm RSD}$. During the later discussion of
reconstruction in this paper, the density field used in reconstruction is
the one that accounts for RSDs.

The convergence, formally written as
\begin{equation}
\kappa=\frac{3}{2}H_{0}^{2}\Omega_{m,0}\int_{0}^{\chi_{\rm CMB}}d\chi\frac{\chi(\chi_{\rm CMB}-\chi)}{\chi_{\rm CMB}}\frac{\delta(\chi)}{a(\chi)}\,,
\end{equation}
can similarly be binned. An expression for convergence binned in discrete
angular pixels that is independent of radial binning is used \cite{0807.3651, 1011.1476},
allowing contributions to be placed into radial bins
that can be summed over later to examine the convergence contribution
from a given radial bin or range of radial bins,
\begin{align}
\kappa_{{\rm bin}} & =\frac{3}{2}H_{0}^{2}\Omega_{m,0}\frac{V_{{\rm sim}}/N_{{\rm sim}}}{\Omega_{bin}}\sum_{{\rm particles,}\,p,\in{\rm bin}}\frac{1}{\chi_{p}a(\chi_{p})}\frac{\chi_{\rm CMB}-\chi_{p}}{\chi_{\rm CMB}}\,,
\end{align}
so that for each angular pixel on the sky the total convergence will be
\begin{equation}
\kappa_{{\rm pix}}=\sum_{{\rm bin}\in{\rm pix}}\kappa_{{\rm bin}}\,.
\end{equation}
We use this convergence map to lens the primary CMB. 

There are several ways to compute the kSZ temperature fluctuations
from particle data. The kSZ temperature fluctuations given by Eq.~\ref{eq1}
can be evaluated by binning the components of the fields $v$
and $\delta$ separately. However, the peculiar velocity field
can be severely undersampled in simulated data, with nonzero velocities
determined by only a single particle, or not at all in some pixels.
In the case of the density field the issue is not as severe, as a
lack of particles is merely indicative of an underdense region, where
the density should be small anyways. A standard practice is therefore
to write the integral in terms of a sum over peculiar particle momenta
\cite{1511.02843},
\begin{equation}
\left(\frac{\Delta T}{T}\right)_{kSZ}=-\frac{\sigma_{T}f_{b}\mu}{\Omega_{{\rm pix}}}\sum_{{\rm particles}\,p\in{\rm bin}}\frac{m_{p}v}{D_{A,p}^{2}}\,.\label{eq:Tksz_binning}
\end{equation}
In standard techniques used to construct kSZ temperature maps, the
only contribution to the temperature field considered is the peculiar
velocity of matter in Newtonian gauge projected along the line of sight,
$v = v_{\rm Doppler, N-body}$. Thus, important contributions
to the observed kSZ temperature perturbations on large angular scales from ISW, SW,
or large-scale velocity modes (modes larger than the simulation volume) have not
been modeled, each of which will contribute to the kSZ temperature fluctuations as
described by Eq.~\ref{eq4}.

In Section~\ref{subsec:Large-scales} we discuss more
precisely how we model these additional contributions, however at the level of
binning, we have two options. We can include these fields at the level
of the already-binned lightcone data, replacing $v$ with
\begin{equation}
\label{eq:veff_contribs}
v \rightarrow v_{\rm eff} = v_{\rm Dopp,\,N-Body} + v_{\rm Dopp,\,LS} + v_{\rm ISW} + v_{\rm SW}
\end{equation}
where the ISW and SW components are given by the respective contributions of
the effects (Eq.~\ref{eq4}) to the temperature perturbation (Eq.~\ref{eq1}), and where
the Doppler contributions are from both the N-body simulation and large-scale (LS) modes
not included in the N-body simulation.
Alternatively, we can compute the
overdensity $\delta_{\rm bin}$ and the velocity $v^{\rm bin}_{\rm eff}$
in each bin, and evaluate Eq.~\ref{eq1} directly. We find that both
methods result in nearly identical kSZ temperature maps and 
power spectra for the angular resolutions we are interested in, although for
the final maps we use Eq.~\ref{eq:Tksz_binning} and~\ref{eq:veff_contribs}.

The final quantity we compute using lightcone data is the peculiar
velocity field, taking the velocity in each bin to simply be the average
velocity of particles within each bin. At low angular resolutions,
which for our simulations means a $\textsc{healpix}$ \cite{astro-ph/0409513} resolution of
Nside=1024, the narrowest redshift-angular bins we consider will typically
contain at least one particle. At higher resolutions, artifacts become
apparent in velocity maps due to undefined velocities in cells without
particles \cite{1312.1022}. However, for the dipole field in particular,
it is sufficient to compare low-Nside maps to our reconstructed
velocity maps as we are interested in reconstructing the dipole 
field on large angular scales ($\ell \lesssim 20$).

As a final point of note, and as a check that the temperature maps and
especially the dipole field reconstruction is insensitive to the
precise binning method used, we employ binning using both a ``nearest
gridpoint'' assignment scheme, and a ``cloud-in-cell''-type
assignment scheme where contributions from individual particles of
fields are distributed to a weighted average of nearby cells, both
radially and in an angular direction. The latter of these methods
introduces additional smoothing, or aliasing, on bin-sized scales;
this suppresses power on these scales, but also suppresses the effects
of shot noise. Despite this difference, we find that the performance of the dipole field
reconstruction presented below is largely insensitive to this detail.

\subsection{Large scales: ``box-in-box''}
\label{subsec:Large-scales}

We formally describe the process of sewing the N-body data and the large-scale modes together using a ``coloring''
operator $\mathcal{C}_{P_c}(f)$ that rescales a stochastic
field $f$ (with its own power spectrum $P_f$) by a power spectrum
$P_c$,
\begin{equation}
\mathcal{C}_{P_c}(f) = \int \frac{d^3 k}{(2\pi)^3} e^{i \vec{k} \vec{x}} f(\vec k) P_c^{1/2}(k)
\end{equation}
so the power spectrum of the resulting field is given by $P_f P_c$.
For a coloring spectrum $P_c = P_f^{-1}$, the field will be whitened.
We additionally make use of an ``inlay'' operator,
$\mathcal{I}(f_1, f_2)$, which acts in configuration space to replace
values in the interior of a (large-scale) field by values of a second (coarsened, small-scale)
field. The procedure of sewing a
small field into another larger field then consists of the following
operations:
\begin{equation}
f_{\rm sewn} = \mathcal{C}_{P_f}( \mathcal{I}( \mathcal{C}_{P_f^{-1}}(f_{\rm lg}), \mathcal{C}_{P_f^{-1}}(f_{\rm sm}) ) )
\end{equation}
Evaluating the above expression entails taking Gaussian random fields
$f_{\rm sm}$ and $f_{\rm lg}$, both with statistical properties
described by $P_f$, whitening these fields, replacing values of the
${\rm lg}$ field by ones from the ${\rm sm}$ field, and finally
de-whitening the fields.

The result of this procedure on the ${\rm lg}$ field is that the
small-scale modes in the region of replacement are now given by modes
from the ${\rm sm}$ field, while large-scale modes have been preserved
and superimposed upon the small-scale field.

Although L-PICOLA provides us with information about the density field,
we are ultimately interested in obtaining the primordial potential, from
which we can compute corrections to the velocity field using linear theory.
In order to obtain the primordial potential on large scales, we extract
the potential on the initial slice using the density field and Poisson
equation\footnotemark,
\begin{equation}
\nabla^2 \Phi = 4\pi G a^2 \delta_\rho^{\rm sim}\,.
\end{equation}
The potential can then be evolved back in time using the transfer function for
the potential, $T(\phi_{\rm sim}\rightarrow\phi_{\rm prim})$.
\footnotetext{As a technical note, we can safely interpret output from the
L-PICOLA simulations in a standard way without worrying about relativistic effects
given our accuracy requirements for the scales we are interested in
\cite{1101.3555,1606.05588,1711.06681}. The evolution of large-scale modes is
determined using linear cosmological perturbation theory, which takes into
account relativistic effects at linear order in metric and density perturbations.}

\begin{figure}[hp!]
  \centering
    \includegraphics[width=0.73\textwidth]{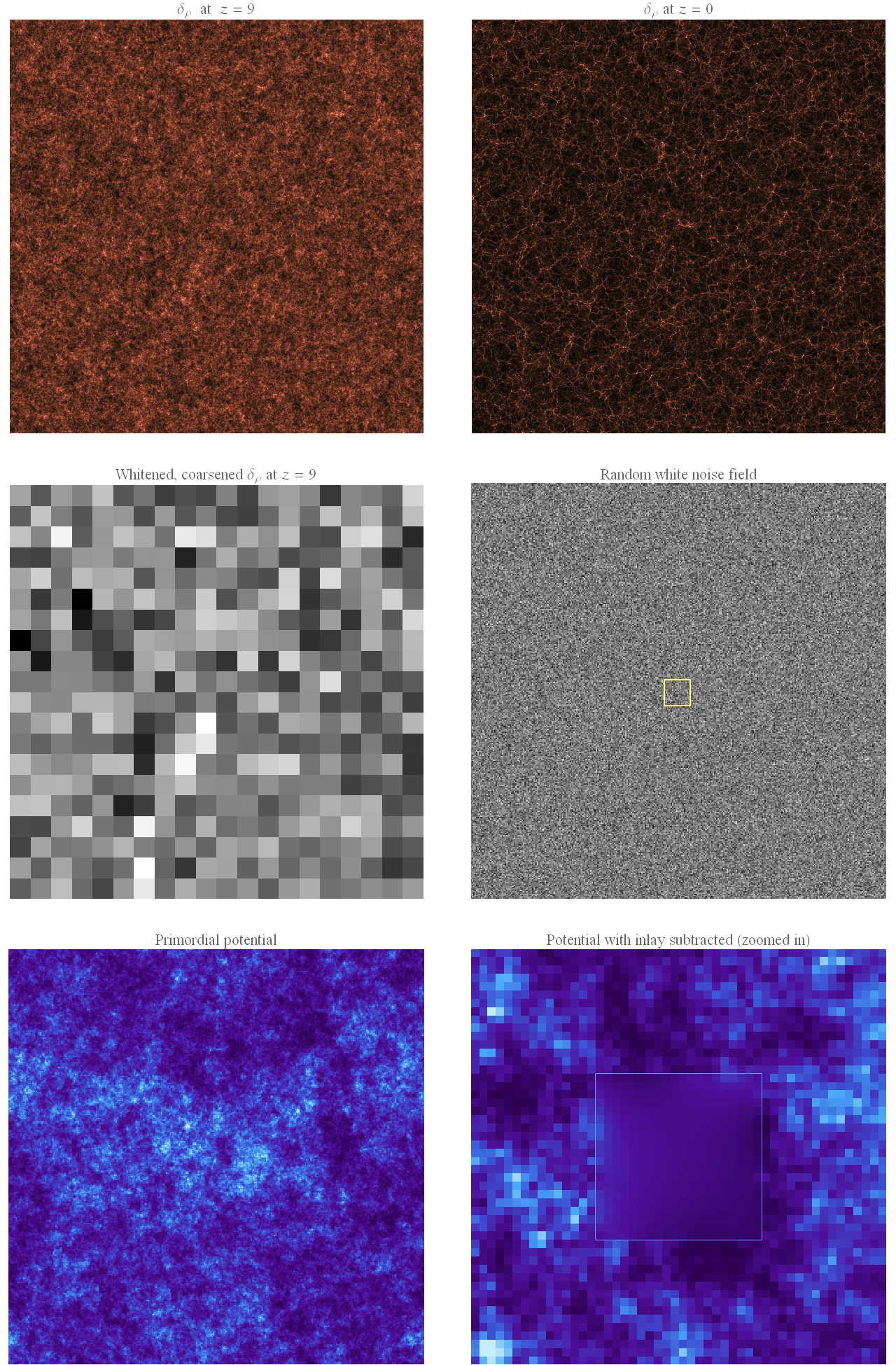}
  \caption{\label{fig:sim_grids} Slices of spatial hypersurfaces of simulations during
  various parts of the sewing-in procedure described by Eq.~\ref{eq:sewing}.
  Top left: the initial L-Picola density contrast field at $z=9$, with comoving box size $L=2\,{\rm Gpc}/h$.
  Top right: the density field at $z=0$.
  Middle left: the initial $z=9$ density field, whitened using the matter power spectrum, averaged over (coarsened)
  so the resolution is the same as that of the box containing large-scale modes.
  Middle right: A random realization of white noise for large-scale modes, with $L=32\,{\rm Gpc}/h$. The central
  $2\,{\rm Gpc}/h$ region that will be replaced has been outlined with a yellow border.
  Bottom left: The primordial potential with white-noise values in the large box replaced using the whitened L-PICOLA field, then colored using the primordial power spectrum.
  Bottom right: The central $5\,{\rm Gpc}/h$, with the colored small-scale box values directly subtracted. Small residual large-scale modes can be seen in the center. The region where the subtraction has been performed is outlined.}
\end{figure}

We are therefore interested in computing
\begin{equation}
\phi_{\rm sewn} = T(\phi_{\rm sim}\rightarrow\phi_{\rm prim})\frac{4\pi G a^2}{\nabla^2} \mathcal{C}_{P_\delta}( \mathcal{I}( \mathcal{C}_{P_\delta^{-1}}(\delta_{\rm Box}), \mathcal{C}_{P_\delta^{-1}}(\delta_{\rm LP}) ) )
\end{equation}
where the L-PICOLA density field is noted by the LP subscript, and the Box subscript
refers to a random realization of a density field with power spectrum $P_\delta$.
Written in Fourier space, the outermost coloring operation, transfer function
operation, and the inverse Laplacian operation can all be combined into an
operation equivalent to coloring by the primordial spectrum. Coloring the large-scale
Box modes with its inverse spectrum is also equivalent to simply generating a field
of white noise, $N_{\rm white}$. Thus the final operation we perform in order to
obtain a large-scale primordial potential consistent with the density field
from the L-Picola simulation is 
\begin{equation}
\label{eq:sewing}
\phi_{\rm sewn} = \mathcal{C}_{P_{\phi,{\rm prim}}}( \mathcal{I}( N_{\rm white}, \mathcal{C}_{P_\delta^{-1}}(\delta_{\rm LP}) )\,.
\end{equation}
The power spectra for the density and primordial potential are both given by the
CLASS code \cite{1104.2933}. We show snapshots of various steps
of this procedure in Fig.~\ref{fig:sim_grids}

Once we have the primordial potential, we use the CMB radiation transfer
functions to obtain the primary CMB, and velocity transfer functions to compute
the contributions to the dipole field due to large-scale modes\footnotemark.
\footnotetext{We could also modulate large-scale modes in the density field, however
long-wavelength density perturbations contribute negligibly to the cross correlation
between the kSZ temperature and density field \cite{1610.06919}, so we do not include
this modulation.} From the large-scale modes, we can then compute the contributions to
$v_{\rm eff}$ from Eq.~\ref{eq:veff_contribs}, and thus their contribution to the
observed kSZ temperature fluctuations.
When computing large-scale contributions to $v_{\rm eff}$, we also need to ensure
we do not double-count modes already accounted for by the N-body simulation.
Therefore, when computing the large-scale Doppler contribution to $v_{\rm eff}$,
we only integrate over modes with wavelengths larger than the N-body simulation
volume.

The CMB multipoles are then computed using the large-volume simulated
primordial potential up to $\ell = 28$. In principle we could generate additional
CMB modes using simulated data, however they will not be correlated with the 
remote dipole field or the density field. We therefore use a random realization of the
primary CMB $a_{lm}$s at $\ell > 28$, based on the theoretical power spectrum obtained from CLASS.

We also include lensing of the primary CMB, utilizing the convergence maps generated
from the lightcone data. From the maps we can compute the lensing potential $\phi$
in harmonic space as
\begin{equation}
\phi_{\ell m} = \frac{2 \kappa_{\ell m}}{\ell (\ell+1)}.
\end{equation}
The lensed CMB temperature is then given by
\begin{equation}
T(\hat{n}) \rightarrow T(\hat{n} + \nabla \phi) \simeq T(\hat{n}) + \nabla \phi \nabla T(\hat{n})\,.
\end{equation} 
Although the CMB is lensed, the kSZ temperature is not. In principle, there could
be lensing of the kSZ temperature fluctuations due to any structures between kSZ
sources and an observer, however we do not model this. The lensing we compute is
also derived from only the N-body volume we simulate, thus in a more realistic treatment,
structure at higher redshifts and on large scales would need to be included. However,
the small-scale density-temperature correlations induced by lensing from the density
field we use for reconstruction are accounted for. In future work, we would
nevertheless prefer to include lensing (and kSZ) contributions from additional
redshifts. 

To model kSZ temperature anisotropies sourced at redshifts beyond our N-body simulation, 
we include Gaussian random noise on angular scales $\ell \agt 1000$ with amplitude 
$\sim 2 \ \mu{\rm K}$. 

\begin{sidewaysfigure}[hp!]
  \centering
    \includegraphics[width=1.0\textwidth]{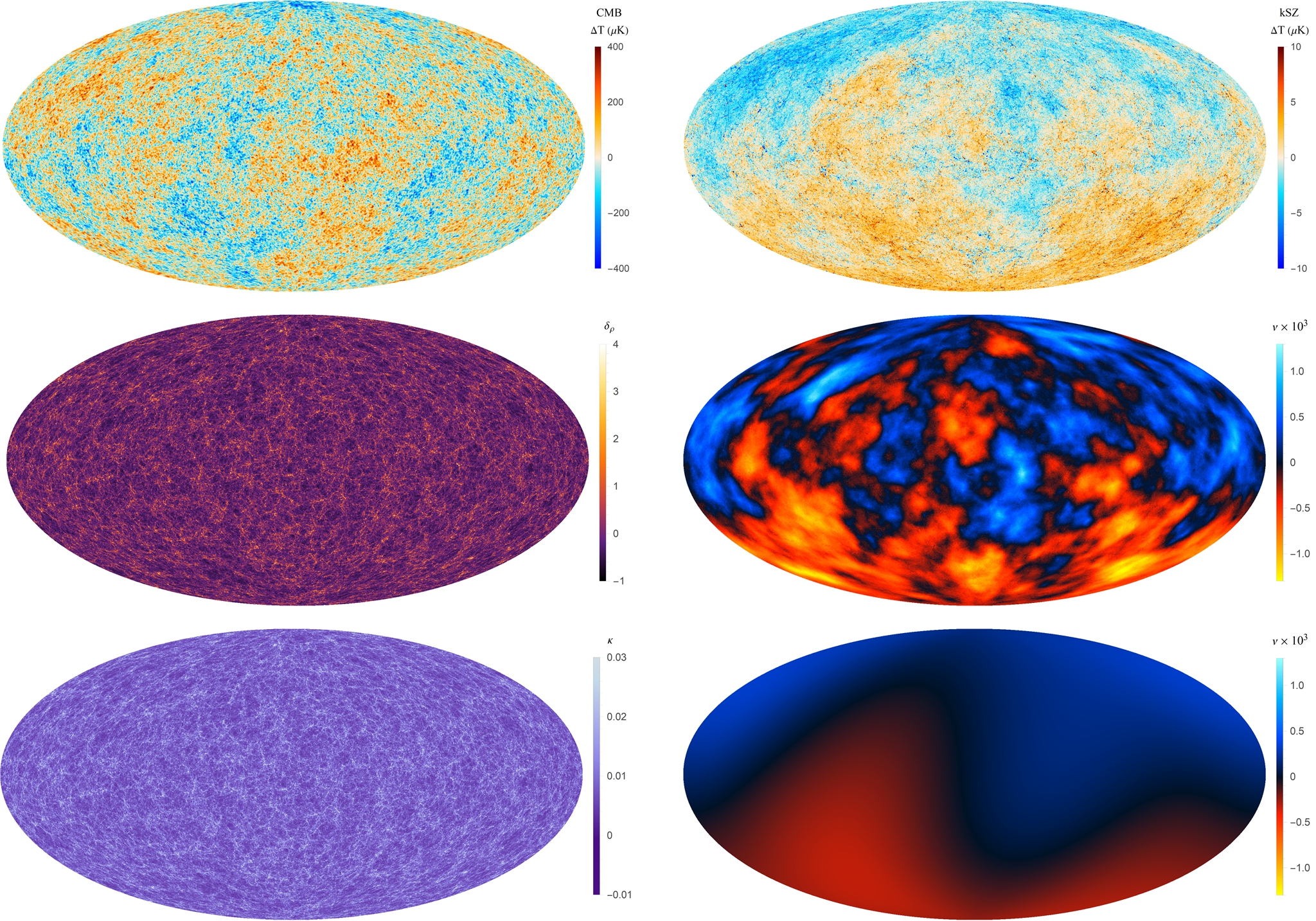}
  \caption{\label{fig:skies} Hammer-Aitoff projections of different fields
  on the sky from the box-in-box simulations; all
  fields are properly correlated.
  Top left: the total CMB temperature fluctuations, including kSZ contributions. The CMB
  dipole is not included.
  Top right: the contribution of the kSZ effect to temperature fluctuations.
  Middle left: the binned, average density field (Eq.~\ref{eq5});
  middle right: the binned, average dipole field;
  bottom left: the binned convergence field;
  bottom right: the contribution to the remote dipole field from the big-box modes.
  Binning is performed over a redshift range $z=0.18$ to $z=0.27$.
  }
\end{sidewaysfigure}

In Figure~\ref{fig:skies}, we show various outputs of our simulation pipeline. Of particular note
is that the that the projected density is properly correlated with the convergence field and the kSZ
temperature anisotropies, and the dipole field is properly correlated with the primary CMB and the kSZ
temperature anisotropies. Focusing on the kSZ map, both the large-scale contributions and large-scale modulation
of power from the dipole field are visible. In addition, the primordial components of the dipole field 
(i.e. contributions from modes in the big box) are visible as responsible for the structure of the dipole field 
on large angular scales.   

\section{Results}\label{sec:results}

\subsection{Reconstruction using a quadratic estimator}
\label{subsec:reconstruction}

We now analyze data from an ensemble of ten simulations to assess 
the performance of the quadratic estimator, Eq.~\ref{eq6}. We utilize two radial binning 
schemes, with the density field on the light cone of each simulation arranged into either
a single bin or eight bins of equal radial comoving width. For each simulation and bin 
we construct maps of $\xi$ defined in Eq.~\ref{eq:xi} and $\zeta^\alpha$ defined in 
Eq.~\ref{eq:zeta}. The power spectra ${C}_\ell^{TT}$, ${C}_{\alpha \ell}^{\delta \delta}$, 
and ${C}^{\delta \tau}_{\alpha \ell}$ used in Eqs.~\ref{eq:xi}, \ref{eq:zeta}, \ref{eq7} to generate the 
$\xi$, $\zeta^\alpha$ fields and reconstruction noise are the sample variances from each realization. We then 
obtain the estimated moments of the binned dipole field from Eq.~\ref{eq6}, and generate a map
of the reconstructed average dipole field in each bin.  

\begin{figure}[htb]
  \centering
    \includegraphics[width=1.0\textwidth]{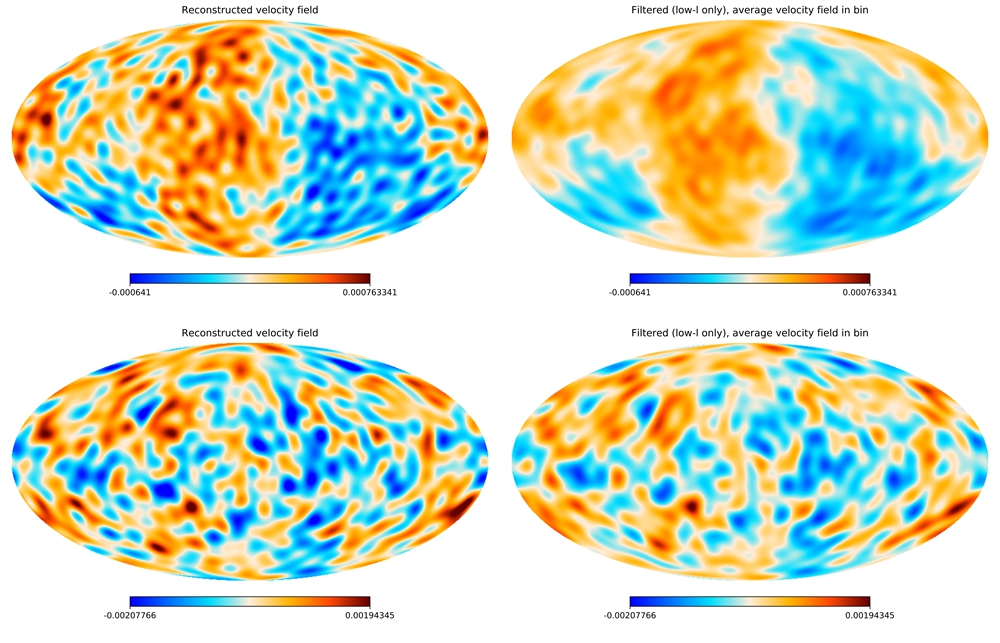}
  \caption{\label{fig:velocity_skies} The remote dipole field obtained from
  simulations compared to the reconstructed remote dipole field. The maps
  do not include modes higher than $\ell > 28$. The reconstruction of
  the top two plots was done using a single redshift bin from
  $z=0.086$ to $z=0.37$, while the bottom plots are a redshift bin from
  $z=0.33$ to $z=0.37$. By eye,
  it is noticeable that large angular modes between the two maps tend
  to agree, while smaller-scale modes only do to a moderate extent. The
  reconstruction of smaller scales is also found to be better in the
  smaller, higher-redshift bin. This is in agreement with
  results obtained by looking at the reconstruction efficiency, shown 
  in the top left panel of Figure~\ref{fig:velocity_reconst}. Excess power
  can also be seen on small scales, consistent with the spectra
  found in Figure~\ref{fig:velocity_ps}.
  }
\end{figure}

In Figure~\ref{fig:velocity_skies} we compare the reconstructed and actual bin-averaged dipole 
fields for a single bin and for the 8th bin of the eight bin configuration. All maps are filtered to contain
only multipoles $\ell < 28$. ``By-eye,'' the reconstruction performs well on large angular scales. We quantify
the agreement between the reconstructed and actual dipole field in two ways. 

\begin{figure}[htb]
  \centering
    \includegraphics[width=1.0\textwidth]{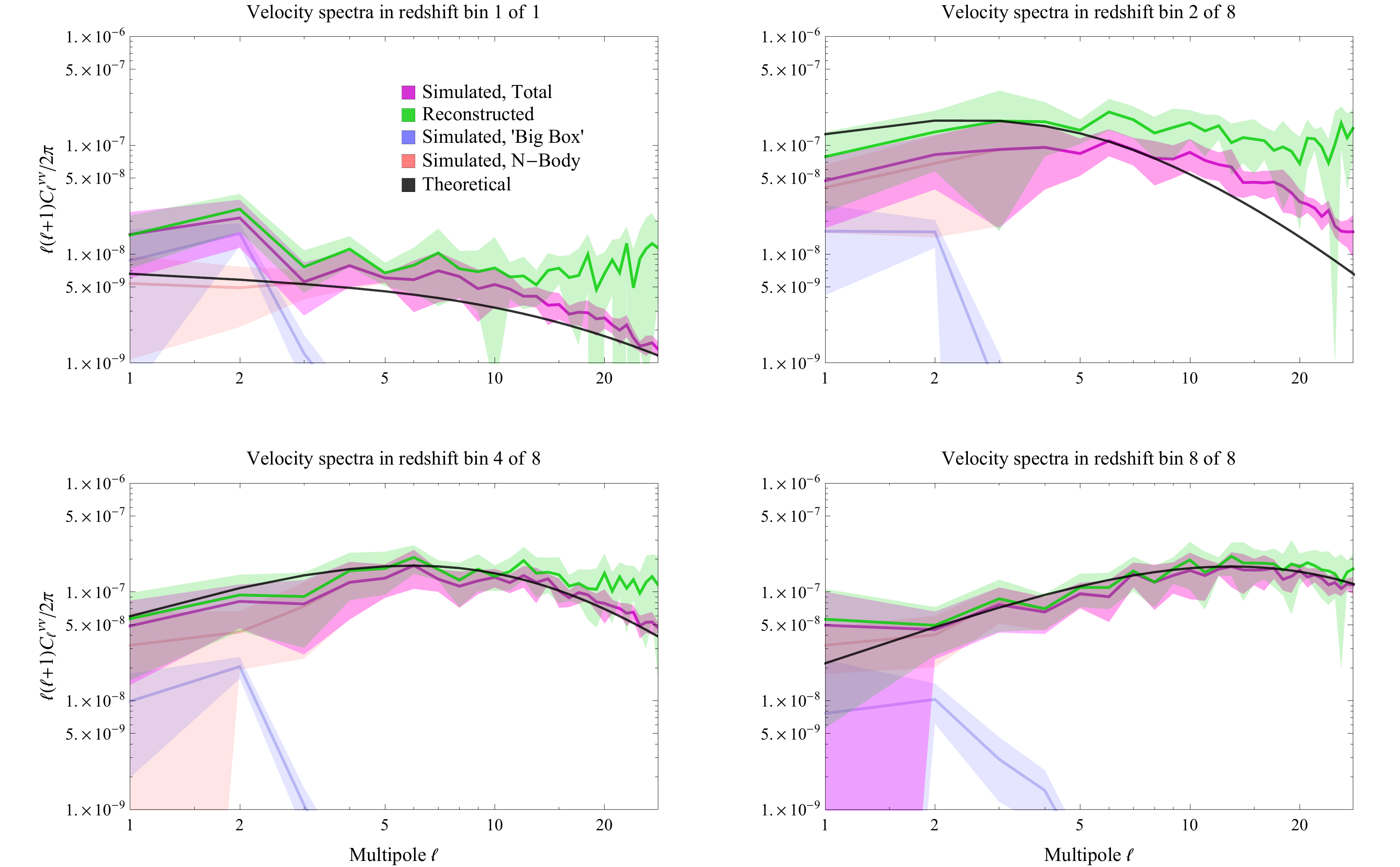}
  \caption{\label{fig:velocity_ps} The velocity power spectra from simulated
  data, compared to the theoretical and reconstructed spectra. Contributions
  to the spectra from the N-body simulation are shown in red, contributions from
  the large-scale box modes in blue, and the total in purple. The reconstructed
  spectra with noise subtracted is in green, and linear theory prediction in black.
  Lines indicate the mean spectrum from our simulations, while solid bands indicate the variance.
  The reconstruction is performed using redshift data in bins over a redshift range
  of $z=0.086$ to $z=0.37$, subdivided into one or eight bins of equal comoving
  distance. Reconstruction efficiencies are shown in Figure~\ref{fig:velocity_reconst}.
  }
\end{figure}

First, we make a comparison at the level of the power spectra in Fig.~\ref{fig:velocity_ps}. 
We compute the mean and standard deviation of the reconstructed dipole field power (with the noise bias removed) 
and the actual dipole field power (total, and separate contributions from the small and big box modes),
as well as the prediction from linear theory using Eq.~\ref{CL_vv}. In this figure, we plot these quantities
for the single bin (top left) and bins 2, 4, and 8 of the eight bin configuration. In general, the agreement between 
the mean reconstructed and the mean actual power is quite good at low multipoles, within a single standard deviation. 
For higher multipoles, the reconstruction is poor and there is an excess of power due to the reconstruction noise.
In addition, there appears to be a systematic bias towards extra power in the reconstructed field at low multipoles,
especially in the single-bin configuration and the lowest redshift bins of the eight bin configuration; the
agreement with linear theory becomes better at higher redshift. This is consistent with a bias due to gravitational non-linearities, 
which we expect to be more important at low redshift. A similar bias exists in
CMB lensing~\cite{1605.01392}, and we hope to investigate this possibility in future work.

As an additional diagnostic of the performance of the reconstruction, we compute the reconstruction efficiency 
\begin{equation}\label{eq:10}
r_L \equiv \frac{\hat{C}_L^{\hat{v} \bar{v}} }{\left(  \hat{C}_L^{\hat{v} \hat{v}} \hat{C}_L^{\bar{v}\bar{ v}}\right)^{1/2}},
\end{equation}
where $\hat{v}$ denotes the reconstructed field and $\bar{v}$ the actual field. The efficiency is not sensitive to an
overall change in normalization, but instead provides us with a measure of how strongly
correlated reconstructed and simulated modes are. In general, we find that the reconstructed
modes agree well with the simulated modes on the largest angular scales. The
reconstruction efficiency is found to be better at higher redshift, again we expect this
due to a lack of nonlinear effects. Reconstruction is also found to perform better in smaller bins, 
an effect we can at least partially attribute to the increased information content: information from 
small-scale modes has not been so heavily averaged away. However, in larger redshift bins, the correlation
with primordial modes is larger, as discussed in the next section.

\begin{figure}[htb]
  \centering
    \includegraphics[width=1.0\textwidth]{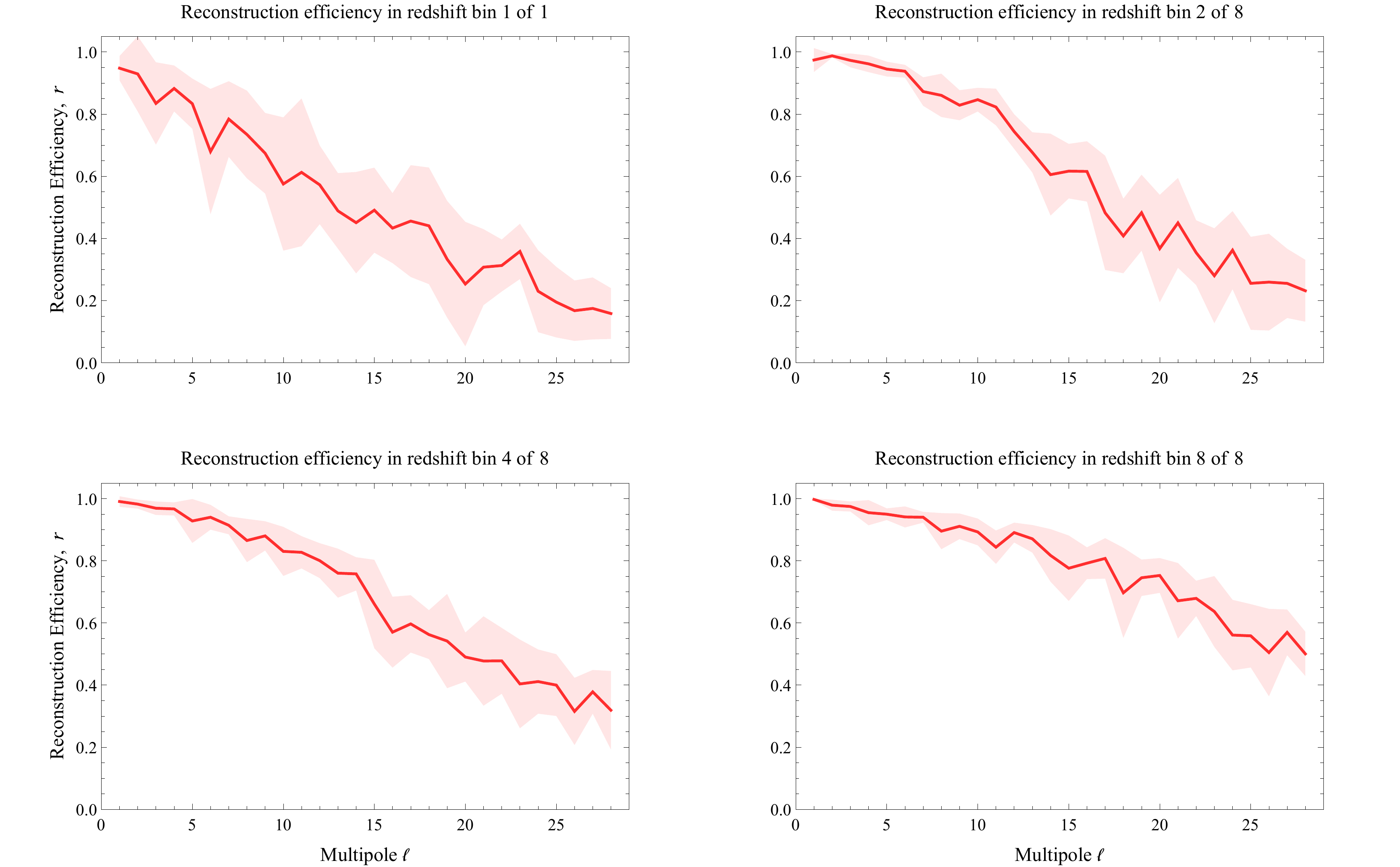}
  \caption{\label{fig:velocity_reconst} The reconstruction efficiency for
  the runs in Figure~\ref{fig:velocity_ps}. The solid line indicates the
  mean reconstruction efficiency in each bin for each simulation realization,
  and the solid band the standard deviation.
  }
\end{figure}

\subsection{CMB-kSZ Dipole Correlation}

We now consider how well we can determine the intrinsic CMB dipole using
information from the reconstructed large-scale velocity field, as suggested 
in Ref.~\cite{1707.08129}.
This idea is not without ambiguity -- because one can arbitrarily change the CMB dipole by 
performing a boost, there is no unique definition of the intrinsic dipole. Instead,  
one must settle on a definition universal and specific enough to facilitate a meaningful
comparison. We can make progress by noting that the local CMB dipole should, to an extent
depending on one's definition of the intrinsic CMB dipole, be correlated with the
$\ell=1$ moments of the remote dipole field. The contributions
to our measured CMB dipole and the remote dipole field of a nearby observer are
determined primarily by small-scale modes which source local peculiar velocities.
However, there are also subdominant contributions to the CMB dipole from larger-scale 
(but still local) velocity modes and other effects both along our past lightcone and at the
CMB last scattering surface.

A standard definition of the fundamental CMB dipole is obtained by boosting to a reference
frame in which the relativistic aberration of the CMB vanishes (see e.g.~\cite{astro-ph/0601594}). 
In Newtonian gauge, this aberration-free dipole is calculated in the frame where an observer 
has vanishing local peculiar velocity, altering the Doppler term in Eq.~\ref{eq4}. A more general
definition of the fundamental CMB dipole is obtained by applying a low-pass filter to the 
Fourier modes contributing to local peculiar velocities. The aberration-free dipole is a special case, 
where all modes contributing to the local Doppler term are filtered out. This more general definition
 is also more closely
related to the dipole field obtained in kSZ tomography, since the bin-averaging 
effectively imposes a low-pass filter on radial peculiar velocities. We will refer to this as the large-scale
Doppler dipole.

We can quantitatively express the correlation between the remote dipole field and
the various definitions of the CMB dipole in terms of transfer functions, with the 
CMB transfer function filtered below a given scale $k_{\rm cut}$,
\begin{equation}\label{CL_Tv}
C_{\alpha 1}^{T\bar{v}} = \int\frac{d^{3}k}{(2\pi)^{3}}P_{\Psi}(k)
{\Delta}^{\bar{v}*}_{\alpha 1}(k) {\Delta}^{T}_{{\rm filt}, 1}(k) \,,
\end{equation}
where as before, $\alpha$ labels a redshift bin in which the remote dipole field $\bar{v}$
is averaged. The filtered CMB transfer function for the dipole is given by
\begin{equation}
\label{eq:filtered_cmb_transfer}
{\Delta}^{T}_{{\rm filt}, 1} = \Theta(k_{\rm cut} - k) {\Delta}^{T}_{{\rm dopp,\,local}, 1}(k)
  + {\Delta}^{T}_{{\rm dopp,\,CMB},1}(k) + {\Delta}^{T}_{{\rm ISW}, 1}(k) + {\Delta}^{T}_{{\rm SW}, 1}(k)
\end{equation}
where $\Theta$ is the Heaviside step function, and the individual contributions to the
radiation transfer function include ISW, SW, and both local and last-scattering-surface
(CMB) Doppler contributions. For the large-scale Doppler dipole, we choose a filtering 
scale equal to the N-body simulation volume ($L_{\rm cut} \sim 2\pi/k_{\rm cut} \sim 3\,{\rm Gpc}$).
For the aberration-free dipole, $k_{\rm cut} \rightarrow \infty$.

\begin{figure}[htb]
  \centering
    \includegraphics[width=0.5\textwidth]{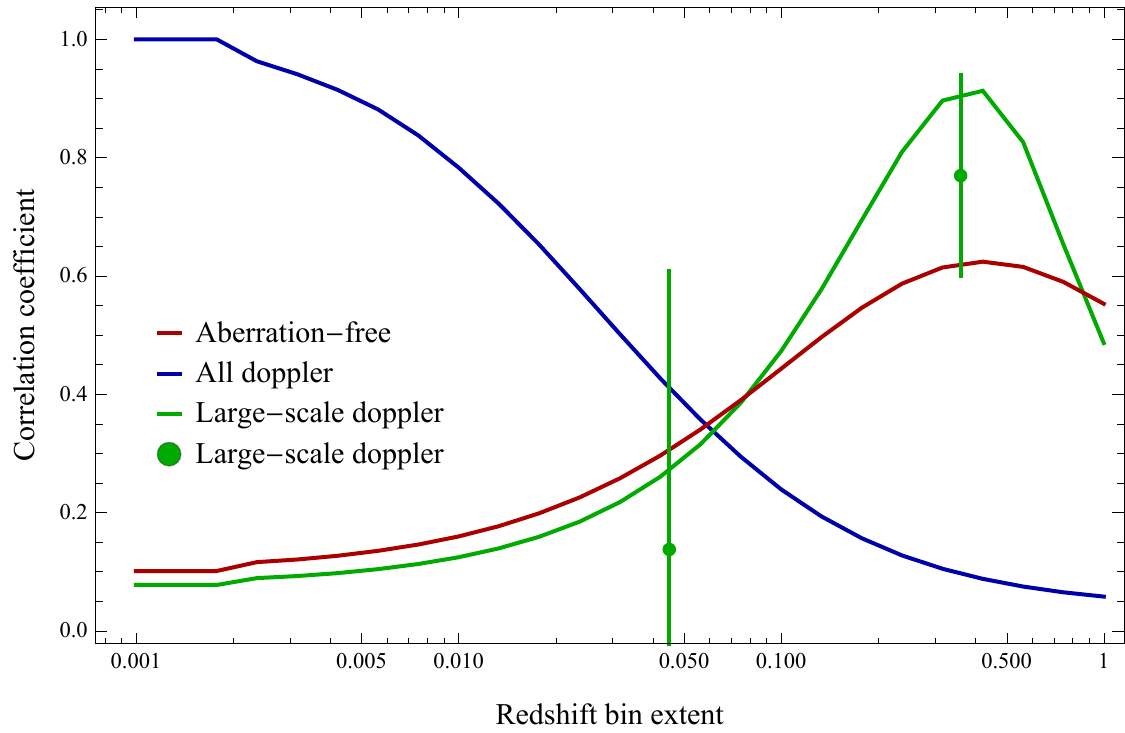}
  \caption{\label{fig:dipole_corr} The reconstructed velocity field and CMB temperature
  dipole correlation coefficient, $C_1^{T\bar{v}}/\sqrt{C_1^{TT}C_1^{\bar{v}\bar{v}}}$,
  computed using different CMB dipoles. The theoretical correlation using the full CMB
  transfer function is shown in blue, correlation with the aberration-free dipole in red,
  and the correlation with ``filtered'' CMB dipole shown in green. Data point show the
  correlation of the simulated CMB dipole filtered on $3\,{\rm Gpc}$ (box-sized)
  scales for two redshift bin sizes. The points are the mean correlation from all
  simulations we perform, and error bars denote the standard deviation.
  }
\end{figure}

In Figure~\ref{fig:dipole_corr}, we plot the theoretical prediction for the correlation 
coefficient (e.g. Eq.~\ref{eq:10}) using linear theory between the $\ell=1$ moment of the bin-averaged 
remote dipole field and three definitions of the CMB dipole: the observed CMB dipole 
(``all Doppler''), the aberration-free dipole, and the large-scale Doppler dipole. We plot the 
theory prediction for a single bin of varying radial extent in redshift. In addition, we show
the mean and standard deviation of the correlation coefficient calculated from ten simulations 
for redshift bins of two different size using the simulated CMB large-scale-filtered dipole and
the reconstructed dipole field. As expected from the discussion above, the correlation between the
observed CMB dipole and the bin-averaged dipole field is small for all but the smallest
bins. Because they are composed primarily of large-scale modes, the correlation between the 
bin-averaged dipole field and the aberration-free and large-scale Doppler dipoles 
improves with bin width. However, the dipole field has a finite correlation length, and
therefore the correlation coefficient eventually goes down. We find that the large-scale 
Doppler dipole can in principle be determined with a maximum correlation coefficient of 
$r \sim 0.9$ while the aberration-free dipole can be determined with a maximum 
correlation coefficient of $r\sim 0.65$. The optimal reconstruction bin width corresponds
to a redshift of $z\sim0.4$. In conclusion, our simulations indicate that 
constraints on the intrinsic CMB dipole should reasonably be attainable in individual realizations.

\section{Discussion and Conclusions}

kSZ tomography is a useful tool for probing the largest
observable scales in our Universe, providing
information in addition to what the primary CMB and large-scale density
surveys alone can tell us. In this work we have explored the ability of a
quadratic estimator to reconstruct the remote dipole field 
using simulated maps of the CMB and density field. 
We have found that the reconstruction process is able to capture
highly significant information about large scales, even
in the presence of physical effects with the potential to contaminate
our ability to reconstruct, including nonlinear growth of structure, RSDs,
lensing, and contributions to the kSZ temperature from structures outside
the range of redshifts considered for reconstruction.

We have accomplished this using a novel simulation technique, in which
a small-scale N-body simulation is sewn into a large-scale volume evolved with
linear theory, allowing us to generate self-consistent maps of kSZ temperature 
fluctuations, the primary
CMB, CMB lensing, density, and dipole fields. In turn, the consistency of these
components allows us to explore the ability of reconstruction techniques
to probe fundamental physics such as determining the intrinsic CMB dipole. Forthcoming
work will additionally allow us to asses the ability of kSZ measurements to
constrain parameters of cosmological models, especially important in the
context of theories competing to describe dark energy and dark matter,
and the presence of unexplained anomalies in the measured CMB.

While this work furthers our confidence in the ability of the reconstruction
procedure to work in practice, it will be important to incorporate additional
physics into our models in order to make future predictions as realistic and robust
as possible. The presence of foregrounds may hamper our ability to reconstruct the
remote dipole field; it will be necessary to ensure we can adequately clean
thermal Sunyaev Zel'dovich emissions, a foreground strongly correlated with the density field. 

We have also relied on several physical assumptions that can potentially
affect the results presented here. One assumption is that the electron
field directly traces the dark matter density field, something that in
practice has been shown to fail on length scales below
$\sim 10$ Mpc \cite{1710.02792}.
Research precisely modeling the impact of physics at play on small scales is
ongoing, however we have found that the angular resolution requirements for
accurate reconstruction of low-$\ell$ modes of the remote dipole field are
fairly mild, thus the physical scales resolved by the simulations are as
well.

Observations will also not provide perfect information about the density
field as assumed in this study; rather, this information is typically
obtained through direct observations, such as of galaxies and clusters at
low redshift, or the 21 cm signal at higher redshifts. Accounting for
these effects in angular maps of the density field can be accomplished in
a future study in which we generate mock galaxy catalogs. 
In principle, existing data of this type can be combined with
the box-in-box technique we introduce in order to produce maps of
increasing realism.

\section{Acknowledgments}

We would like to thank Neal Dalal, Moritz Munchmeyer and Marcel Schmittfull for helpful discussions.
We would also like to thank Tom Giblin for discussions and for his help in
accessing computational resources. The simulations in this work made use of
hardware provided by the National Science Foundation and the Kenyon College
Department of Physics. This research was supported in part by Perimeter Institute for Theoretical Physics. 
Research at Perimeter Institute is supported by the Government of Canada through the Department of Innovation, 
Science and Economic Development Canada and by the Province of Ontario through the Ministry of Research, 
Innovation and Science. MCJ was supported by the National Science and Engineering Research Council 
through a Discovery grant. JBM acknowledges support as a CITA National Fellow.

\bibliography{references}

\end{document}